\documentclass[twocolumn]{aastex61}
\usepackage{ulem}
\usepackage{xcolor}
\usepackage{graphicx}
\usepackage{enumerate}

\shorttitle{Supernova Remnants and their Circumstellar Environments}

\shortauthors{Patnaude et al.}

\begin{document}

\title{The Impact of Progenitor Mass Loss on the Dynamical and Spectral 
Evolution of Supernova Remnants}

\correspondingauthor{Daniel J.~Patnaude}
\email{dpatnaude@cfa.harvard.edu, herman@kusastro.kyoto-u.ac.jp}

\author{Daniel J.~Patnaude}
\affiliation{Smithsonian Astrophysical Observatory, Cambridge, 
MA 02138, USA}

\author{Shiu-Hang Lee}
\affiliation{Kyoto University, Department of Astronomy,
Oiwake-cho, Kitashirakawa, Sakyo-ku, Kyoto 606-8502, JP}

\author{Patrick O.~Slane}
\affiliation{Smithsonian Astrophysical Observatory, Cambridge, 
MA 02138, USA}

\author{Carles Badenes}
\affiliation{Department of Physics and Astronomy and Pittsburgh
Particle Physics, Astrophysics and Cosmology Center (PITT PACC), 
University of
Pittsburgh, 3941 O'Hara St, Pittsburgh, PA 15260, USA}
\affiliation{Institut de Ciencies del Cosmos, Universitat de Barcelona
(ICCUB), Marti i Franques 1, Barcelona, 08028, Spain}

\author{Shigehiro Nagataki}
\affiliation{RIKEN, Astrophysical Big Bang Laboratory \& Interdisciplinary 
Theoretical and Mathematical Science Program, 
2-1 Hirosawa, Wako, Saitama 351-0198, JP}

\author{Donald C.~Ellison}
\affiliation{Department of Physics, North Carolina State University,
Raleigh, NC 27695-8202, USA}

\author{Dan Milisavljevic}
\affiliation{Smithsonian Astrophysical Observatory, Cambridge, 
MA 02138, USA}
\affiliation{Purdue University, Department of Physics and Astronomy, 525 Northwestern Avenue, West Lafayette, IN 47907}

\begin{abstract}

There is now substantial evidence that the progenitors of some
core-collapse supernovae undergo enhanced or extreme mass loss
prior to explosion. The imprint of this mass loss is observed
in the spectra and dynamics of the expanding blastwave on timescales
of days to years after core-collapse, and the effects on the spectral
and dynamical evolution may linger
long after the supernova has evolved into the remnant stage. In this
paper, we present for the first time, largely self-consistent 
end-to-end simulations for the evolution of
a massive star from the pre-main sequence, up to and through 
core collapse, and into the remnant phase. We present three models
and compare and contrast how the progenitor mass loss history
impacts the dynamics and spectral evolution of the supernovae
and supernova remnants. We study a model which only includes steady
mass loss, a model with enhanced mass loss over a period of $\sim$
5000 years prior to core-collapse, and a model with extreme mass
loss over a period of $\sim$ 500 years prior to core collapse. The 
models are not meant to address any particular supernova or supernova
remnant, but rather to highlight the important role that the
progenitor evolution plays in the observable qualities of supernovae
and supernova remnants.
Through comparisons of these three different progenitor evolution
scenarios, we find that the mass loss in late stages (during and after core 
carbon burning) can have a profound impact on the dynamics and 
spectral evolution of the supernova remnant centuries after
core-collapse.

\end{abstract}


\section{Introduction}

When interpreting the remnants of core collapse supernovae (CCSNe), 
assumptions regarding the isotropy of the progenitor mass loss
are frequently made. Often times, it is assumed that the mass loss
remained steady up to core collapse \citep[c.f.,][]{chevalier05}. 
However, the endpoint in massive star evolution is poorly 
understood. In particular, violent and episodic mass 
loss is now observed in the progenitors of some core
collapse supernovae (CCSNe), most notably in 
SN~2009ip \citep[$\dot{\mathrm{M}}$ $\sim$ 0.1 M$_{\sun}$ yr$^{-1}$;][]
{pastorello13,mauerhan13,fraser13,margutti14,smith14c},
though substantial evidence exists for extreme mass loss
in other Type IIn supernovae \citep{ofek13,ofek14,elias16,smith17}.
More exotic mass loss is ascribed to Type Ibn SNe 
\citep[e.g., SN~2006jc;][]{pastorello08} and Type Ib/c SN that 
transition to Type IIn after some time, including SN~2001em \citep{chugai06},
and SN~2014C \citep{milisavljevic15,margutti17}. 
Even in ``normal'' SN IIb/IIL and
SN IIP, evidence for enhanced mass loss \citep[relative to rates observed
in red supergiants;][]{smith14a} is observed 
\citep[e.g.,][]{milisavljevic13,maeda15,kamble16,chakraborti16,morozova17}.

The origin of the extreme mass loss remains a mystery, but
several theories have been suggested. 
For instance, the onset of core carbon and oxygen
burning can lead to stellar cores that are super-Eddington. Some of 
this energy could be tapped
by as of yet poorly understood processes in the core 
\citep{quataert12,shiode14}. 
If the progenitor is sufficiently compact, convectively
driven waves could rise to the surface and unbind
envelope material, depositing it into the 
circumstellar environment. Similarly, nuclear 
shell burning could lead to unstable flows near the
surface, possibly also ejecting material on 
timescales that are short relative to the life
of the progenitor \citep{smith14b}. 

On longer timescales, stable 
mass transfer or common envelope evolution could
remove material from the surface of the progenitor \citep{demink13}.
In the case of a common envelope binary system, the H-rich
envelope could be removed prior to core collapse
\citep[e.g.,][]{podsiadlowski92}. Whatever the
mechanism, evidence for enhanced mass loss prior
to core collapse is observed, either directly through massive
eruptions by so called supernova-impostors, or
through the interaction of the blastwave with
a circumstellar shell of ejected material, as in some IIn
or even more typical IIP/IIL/IIb supernovae. 

\begin{figure}
\includegraphics[width=0.5\textwidth]{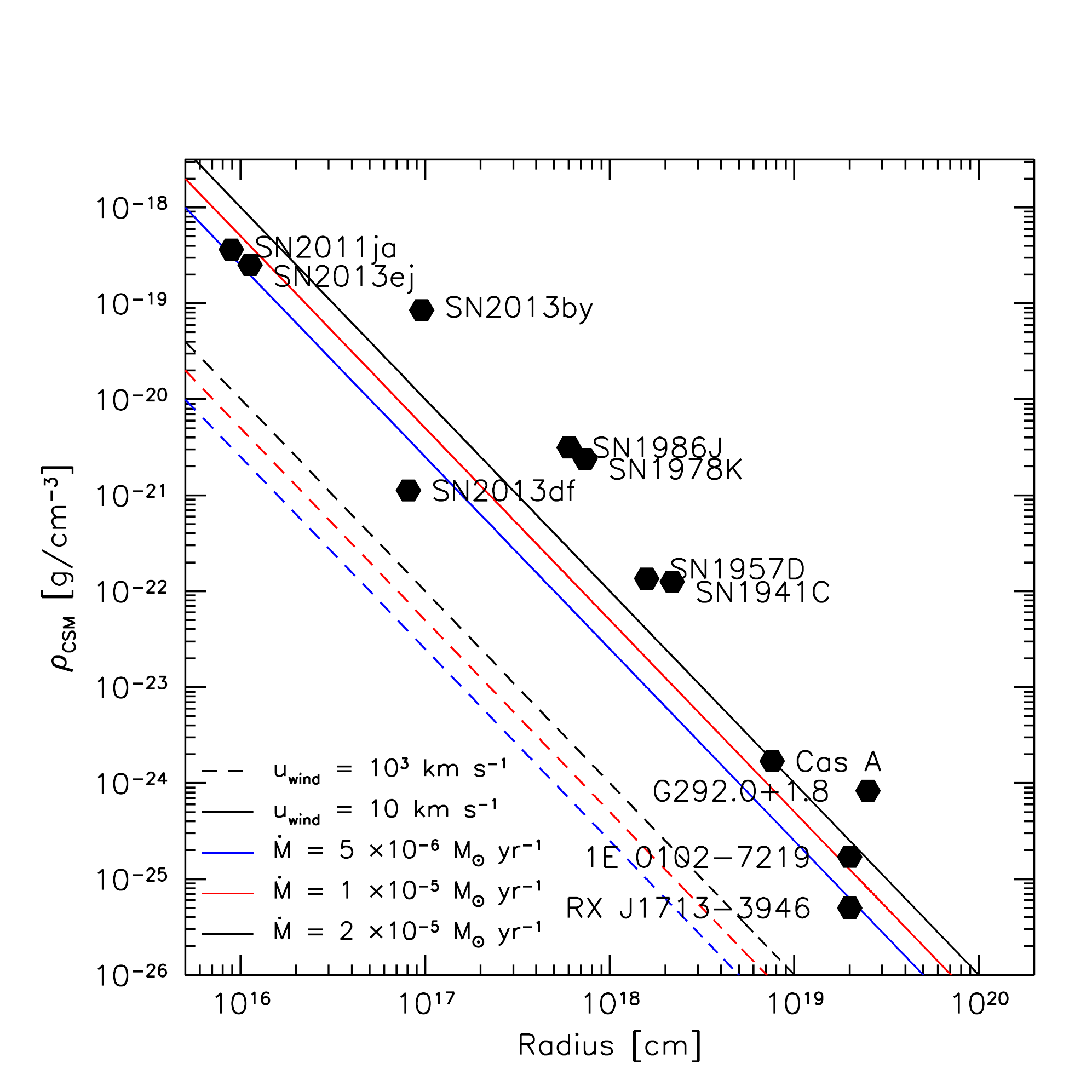}
\caption{Comparison of SNe and SNR radii with those predicted via various
mass loss rates and outflow speeds. The SNe/SNR radii can be related back
to the time before core collapse, if assumptions about the shock speed
and wind speed are made, since $t_{\mathrm{shock}}$ = $v_{\mathrm{wind}} 
\times t_{\mathrm{wind}}/v_{\mathrm{shock}}$. For supernovae, radii 
are derived from the X-ray emission, assuming a wind speed of 10 km s$^{-1}$
\citep{immler03}. Data are from \citet{ross17,long12,soria08}. Data
for supernova remnants are from \citet{patnaude09}, \citet{lee10} and
\citet{ellison12}.}
\label{fig:cavity}
\end{figure}

As illustrated in Figure~\ref{fig:cavity}, while supernovae 
sample the stages of evolution much
closer to core collapse, supernova remnants (SNRs) typically
probe stellar evolution on much longer timescales \citep[c.f.,][]{patnaude17}.
This is because the timescale for the evolution of the
circumstellar environment is set by the outflow speed
of the wind, while the evolutionary timescale for the
supernova remnant is determined by the blastwave velocity.
For example, a 10000 km s$^{-1}$ shock that expands into
a 10 km s$^{-1}$ wind samples 1000 years of stellar evolution for
every year of blastwave evolution -- a 100 year old SNR has probed a 
significant portion of the red supergiant (RSG) phase of a massive star's
life. This would imply that SNR shocks are not effective probes of
the latter stages of stellar evolution, since they
interact with material lost primarily during core helium 
burning.

However, while the characteristic timescale for SNR evolution is 
a few thousand years and is dependent upon the
explosion energetics, ejecta mass, and circumstellar density
\citep{truelove99}, the timescale for ions and electrons to recombine in a
partially ionized plasma is $\sim$ 10$^{12}$/$n_e$ s, where $n_e$ is
the number density of electrons. The circumstellar density
around a massive star is generally thought to decrease with
increasing radius due to flux conservation in the stellar wind
\citep{dwarkadas05,dwarkadas07,dwarkadas12}. 
So, as the supernova shock expands into the wind, the density of shocked 
material decreases, so that $n_e$ is a decreasing function of SN age. This
suggests that the recombination timescale in the plasma increases with
increasing remnant age, and any circumstellar interaction that occurs
early in the remnant evolution could be detectable at later remnant
ages.

Often times, the X-ray spectrum from shocked circumstellar material
is modeled as a blastwave interacting with a progenitor wind where the 
density $\rho_{\mathrm{CSM}}$ $\propto$ 
$r^{-2}$. However, in \citet{patnaude15}, we demonstrated that many SNRs
do not always show X-ray emission that is consistent with
the interaction between the ejecta and a power-law wind.
We argued that enhanced mass loss prior to core collapse 
could greatly increase the X-ray luminosity without strongly
affecting the blastwave radius. This is because the X-ray emission scales like
the density squared, while the blastwave radius is only a weak 
function of the ambient medium. 

In this paper, we extend \citet{patnaude15} by following
the end-to-end evolution of a massive star, from
the pre-main sequence, through the remnant phase. We 
accomplish this by using the {\tt MESA} stellar evolution
code to construct three different massive progenitor
scenarios -- one where the progenitor loses very little
mass to a standard power-law wind; one where the star
is stripped of some of its envelope during core carbon
burning; and one where the H-envelope is almost entirely removed 
during core oxygen burning. We use the mass loss history of
the progenitor to construct its circumstellar environment.  
Using a version
of {\tt SNEC} modified to follow the explosive nucleosynthesis
that occurs during core collapse, we explode
these progenitors, computing the velocity and density
fields of the ejecta, as well as its composition. 
We follow the evolution of these
models to ages of 400 year with our {\tt ChN} code, simulating
CSM properties appropriate for each mass loss
scenario.
In Section~\ref{sec:e2e}, we describe in detail each model
component, and address the uncertainties associated
with each model. In Section~\ref{sec:results} , we present and discuss 
our models. We present our conclusions in Section 4.

\section{End-to-End Models}
\label{sec:e2e}

In this section, we discuss the chain of models we use to 
simulate the stellar evolution up to, through, and beyond
core collapse. In each subsection, we discuss the models used for
each evolutionary stage. In the last subsection we discuss 
the uncertainties in our approach. We stress that no model
is tailored to address any particular supernova or remnant, and
that we are presenting a parametrized framework with which we
can study more specific scenarios in the future. 

\subsection{Stellar Evolution Models}
\label{sec:mesa}

Models for 15M$_{\sun}$ progenitors are evolved using
Modules for Experiments in Stellar Astrophysics 
\citep[hereafter {\tt MESA}, version 8845;][]{paxton11,paxton13,paxton15}\footnote[1]{http://sourceforge.mesa.net}. 
All models are initialized at solar metallicity ($Z = 0.02$). We evolve 
three models, without rotation, but we
employ the ``Dutch'' wind-scheme \citep{nieuwenhuijzen90,nugis00,
vink01,glebbeek09} with an efficiency $\eta = 0.8$. Each model
is evolved from the pre-MS through core collapse, which we define
as the time when the infall velocity at any location is $\geq$ 1000
km s$^{-1}$. For each model, we use {\tt MESA} inlists made
available from \citet{farmer16}. We follow their scheme for specifying the
mass and temporal evolution of the models during burning phases
leading up to core collapse.

We make use of the {\tt aprox21} nuclear burning network. This
is chosen for (1) speed and efficiency, and (2) to match the burning
network we have added to our supernova models (Section~\ref{sec:snec}). 
While \citet{farmer16} concluded that the final electron
fraction and mass locations of the primary nuclear burning shells can 
vary by $\approx$ 30\% based on the choice of nuclear burning network,
and that a minimum of 127 isotopes are needed in order to gain
convergence in these values at levels of 10\% or better, we note that
we are interested
in the bulk qualities of X-ray spectra from astrophysically abundant
elements (O, Si, S, Fe, etc.). Additionally, since 
we are comparing the synthesized
X-ray spectra amongst models and not making any comparisons of final
elemental yields to observations -- we feel that our choice of nuclear
burning network satisfies our requirement for speed and efficiency while
also capturing the spirit of the necessary physics.

We specify three models. Model {\tt m15Iso} uses the Dutch wind
models up to the onset of core collapse. The average mass loss rate
over the lifetime of the star is $\sim$ 5$\times$ 10$^{-6}$ 
M$_{\sun}$ yr$^{-1}$. Model {\tt m15C} includes
enhanced mass loss during core carbon burning, at a constant
rate of 10$^{-4}$ M$_{\sun}$ yr$^{-1}$, resulting in $\sim$ 2 additional
solar masses of material deposited into the circumstellar 
environment. When core oxygen
burning begins, we revert to mass loss rates from the Dutch scheme. Model
{\tt m15O} follows isotropic mass loss up to core oxygen burning,
at which point we employ extreme mass loss. In this case, the mass
loss is held fixed at 0.1 M$_{\sun}$ yr$^{-1}$. We include this extreme
mass loss until the formation of the silicon core. During this
phase, the progenitor loses $\sim$ 6M$_{\sun}$ of material. At the
onset of core collapse, our models have final masses of 
$\sim$ 13, 10, and 6M$_{\sun}$. 

We stress that the enhanced and extreme mass loss models are not meant
to represent any physical processes associated with late stage stellar 
evolution. While there is growing evidence for enhanced mass loss in
supernova progenitors prior to core collapse, both on timescales of
$\sim$ a few thousand years, down to timescales of a few years, the
mechanism for this mass loss remains poorly understood, and could
result from hydrodynamic instabilities and turbulence in the outer
layers of the star \citep{smith14b}, wave-driven mass loss due to 
energy extraction from the super-Eddington core 
\citep{quataert12,shiode14}, or pulsational-driven 
superwinds \citep{yoon10} amongst other possibilities. In this study,
we are aiming to deposit the mass in the CSM at a position that is
self-consistent with the timing of the mass loss epoch.
%One thing is 
%clear however -- enhancing the mass loss during late stage stellar
%evolution has the effect of prolonging the burning phase that the
%evolution is occuring in.

\begin{figure}
\includegraphics[width=0.5\textwidth,viewport=210 10 571 560]{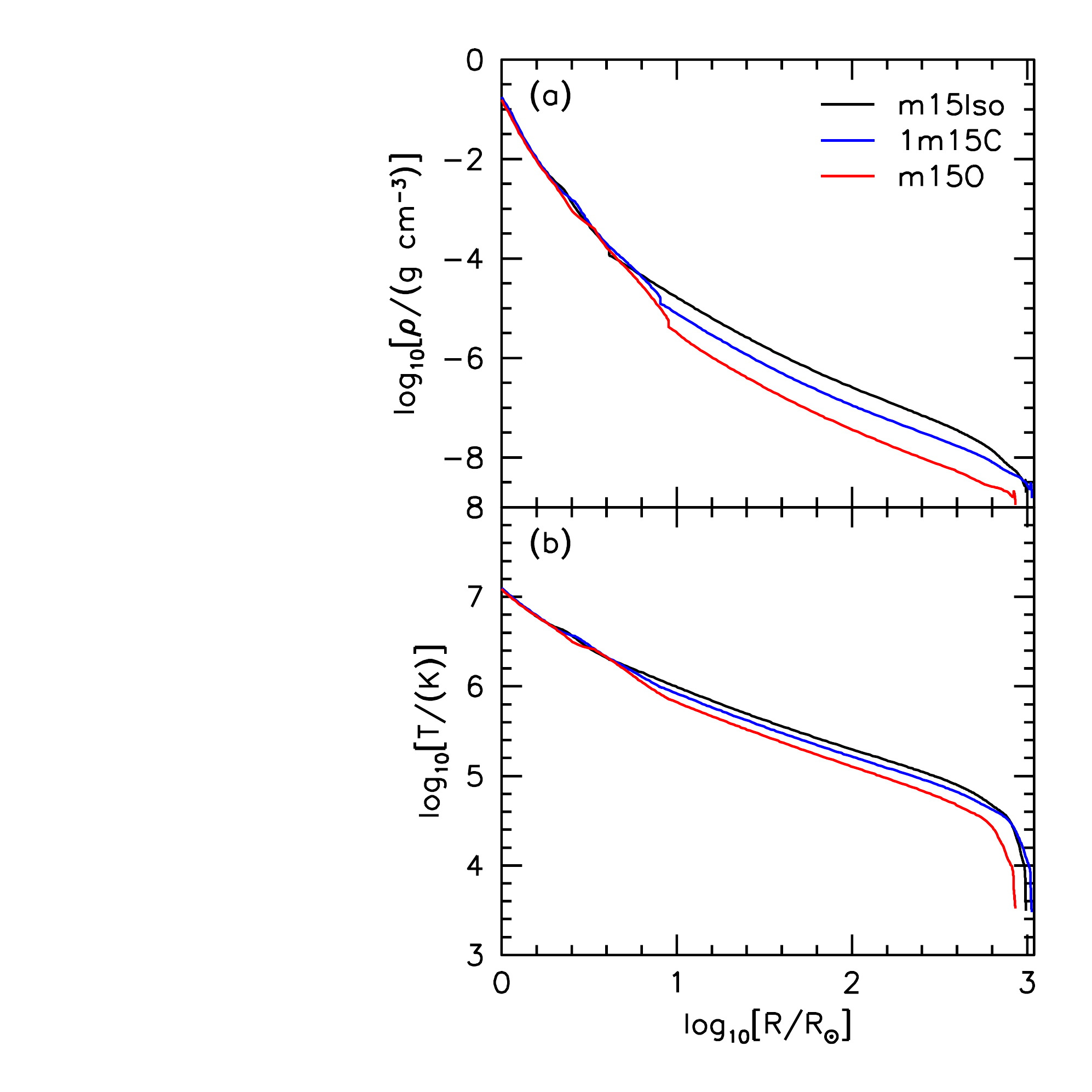}
\caption{{\it Upper}: Density as a function of stellar radius for each
model at the onset of core collapse. {\it Lower}: Final temperature
for each model.}
\label{fig:star}
\end{figure}

While the
exact mass loss mechanism will undoubtedly alter the final evolution
of the progenitor, we look to understand how the mass loss affects the
evolution of the SNR, independent of how it arose. In Table~\ref{tab:mesa}, 
we present the initial and final parameters for each model, while in
Figure~\ref{fig:star}, we plot the density and temperature of the
progenitors for each model at the onset of core collapse. 
Beyond the large differences in final mass, brought about by the
choice of fiducial mass loss, there are not large differences in the
final parameters of the progenitors, with the most notable differences
being in the final progenitor radii. As seen in Figure~\ref{fig:star},
the thermodynamic profile of the progenitor interior of $\approx$ 6R$_{\sun}$
are virtually identical. However, since nuclear reaction rates 
are sensitive to changes in temperature and density, it is these small
differences that can lead to differences in the final compositions. Likewise,
these small differences could lead to larger differences in the
structure of the ejecta. For instance, the amount of envelope retained
by the progenitor prior to core collapse could affect the growth 
of instabilities during the explosion \citep[c.f.,][]{wongwathanarat15}.

\subsection{Supernova Models}
\label{sec:snec}

The {\tt MESA} models are coupled to a spherically symmetric 
(1D) Lagrangian hydrodynamics code that uses equilibrium-diffusion
radiation transport. The code, called the SuperNova Explosion
Code (hereafter referred to as 
{\tt SNEC}\footnote[2]{http://stellarcollapse.org/SNEC}) is made
freely available, and in its publicly available form, follows the
time dependent radiation hydrodynamics and other basic physics 
needed for supernova light curve generation and $^{56}$Ni heating. 
A detailed discussion of {\tt SNEC} may be found in 
\citet{morozova15,piro16,morozova17}. 

In its basic form, {\tt SNEC} couples to a model for the 
structure and composition of the supernova progenitor. It includes the
ability for an arbitrary composition, and includes a prescription
for mixing of ejecta, via a boxcar smoothing algorithm. Observations
of supernovae and remnants show evidence for both mixing of Ni-rich ejecta
during the explosion \citep[e.g., SN~1987A;][]{li93}, as well as
evidence for Rayleigh-Taylor mixing between layers of differing 
composition and densities, and these effects are confirmed in multidimensional
numerical modeling of core collapse SNe during the first few seconds
of evolution \citep{janka12,wongwathanarat15,wongwathanarat17}. 
While an approximation for the 
Rayleigh Taylor instability exists in one dimension \citep{duffell16}, 
for this study we choose to ignore
the mixing of metal-rich ejecta into the outer layers of the
progenitor. As substantial evidence exists for the mixing of Ni and Fe-peak
elements into the outer layers of ejecta, in both supernovae and some
supernova remnants, we will explore mixing in followup papers.

\begin{figure}
\includegraphics[width=0.5\textwidth,viewport=206 10 550 554]{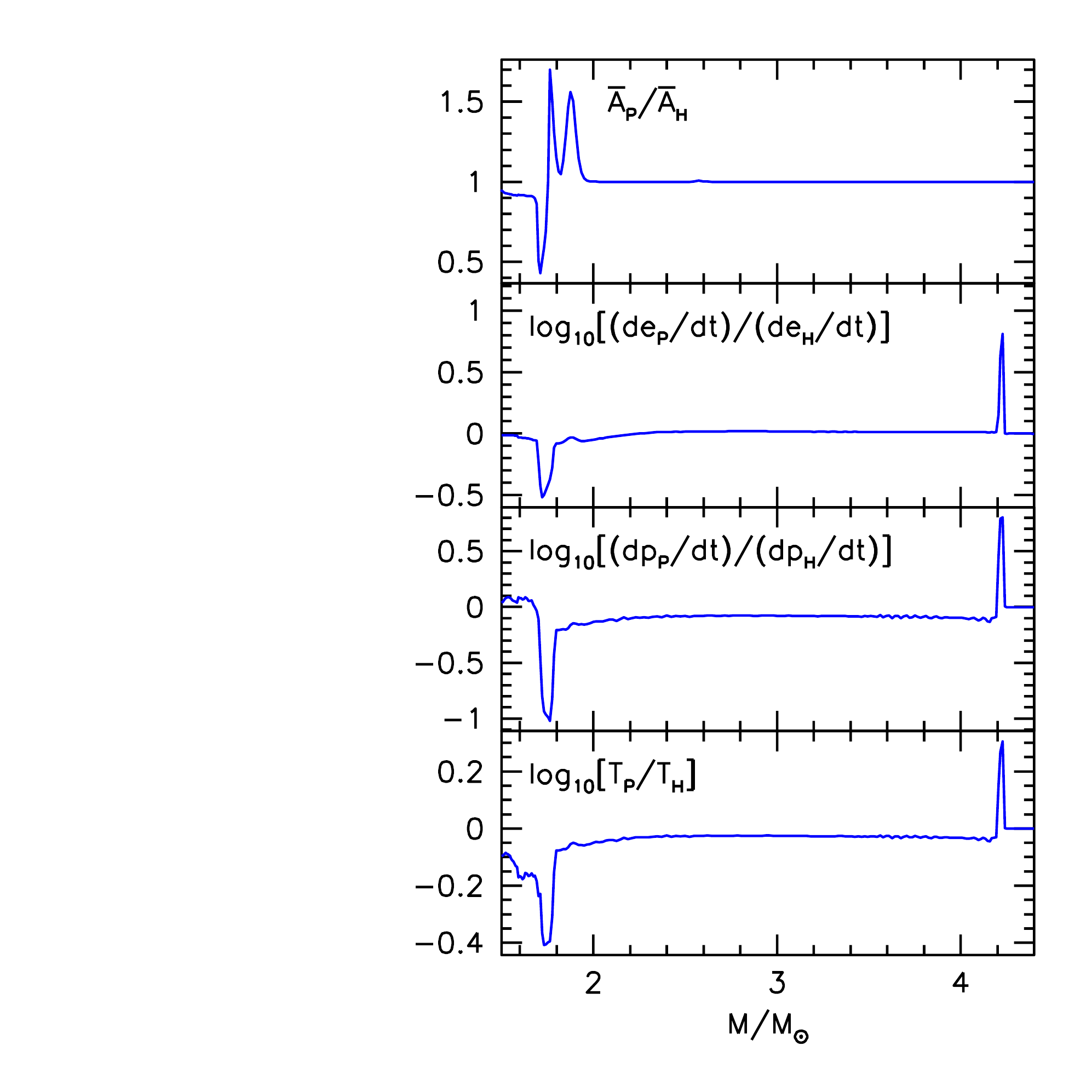}
\caption{Differences between the Paczynski and Helmholtz Equations of
State. The top panel shows differences in $\bar{A}$ (the mean atomic mass), 
the middle two panels
show differences in the time derivatives of internal energy and pressure, 
and the bottom panel shows differences in the temperature.}
\label{fig:eos}
\end{figure}

{\tt SNEC} allows the user to choose between either a thermal bomb
or piston driven explosion. We choose to use the thermal bomb
method, whereby we specify the core mass to be excised, and then energy
is deposited in a user-specified number of mass zones, at which point
the explosion calculation begins. This is a completely {\it ad-hoc} method 
and ignores the important fact that the explosion is likely driven 
by a combination of neutrino heating of the shock and hydrodynamical
instabilities. 

Finally, {\tt SNEC} closes the system of hydrodynamic conservation laws with 
the choice of an equation of state (EOS). The choices are either an ideal
gas, or the Paczynski EOS \citep{paczynski83}, which includes 
contributions to the total pressure from radiation, ions, and
electrons. While the Paczynski EOS may provide a rough
approximation, it is not thermodynamically consistent, does not
treat pair-production, does not use a chemical potential, and is not
suitable for an arbitrary composition
\footnote[3]{Timmes; private communication}. Based on these limitations, we
chose instead to incorporate the Helmholtz EOS 
\citep{timmes00}\footnote[4]{http://cococubed.asu.edu/code\_pages/eos.shtml}.
The Helmholtz EOS is thermodynamically consistent. In Figure~\ref{fig:eos},
we plot thermodynamic quantities and their derivatives as a function
of mass coordinate in {\tt SNEC}. The calculations shown in
Figure~\ref{fig:eos} are for after explosive nucleosynthesis ceases. 
The largest differences, of order 0.5 dex, 
occur at both the location of the shock, and
in areas just above the explosion launch point. Differences between the
two EOS are most readily seen as small differences in the temperature of the
shocked material, as well as differences in the pressure derivative. 
Nuclear burning is sensitive to the temperature, so anomalous heating
due to an imprecise EOS can lead to extra burning in the ejecta. This is
seen as a difference in $\bar{A}$ in the top panel of Figure~\ref{fig:eos}. 

As mentioned, {\tt SNEC} allows for arbitrary composition, but it does
not include a way to update the composition due to explosive burning. 
Certainly, the additional nucleosynthesis during the explosion will
not impact the observable properties of swept up material from the
CSM interaction, but it could alter the measured abundances in the 
ejecta. We have chosen to implement nuclear burning by using the
{\tt aprox21} 
network\footnote[5]{http://cococubed.asu.edu/code\_pages/burn.shtml}. This
is the same network we use in our {\tt MESA} models, and is an
adequate, yet incomplete burning network. For efficiency, we follow the
burning in each mass shell until the temperature falls below 
10$^{7}$ K. As discussed in \citet{farmer16}, the final composition is
sensitive to the nuclear reaction network chosen. We defer a study 
of larger networks to subsequent papers.

\begin{figure*}
\begin{minipage}[c]{1.\textwidth}
\includegraphics[width=0.5\textwidth]{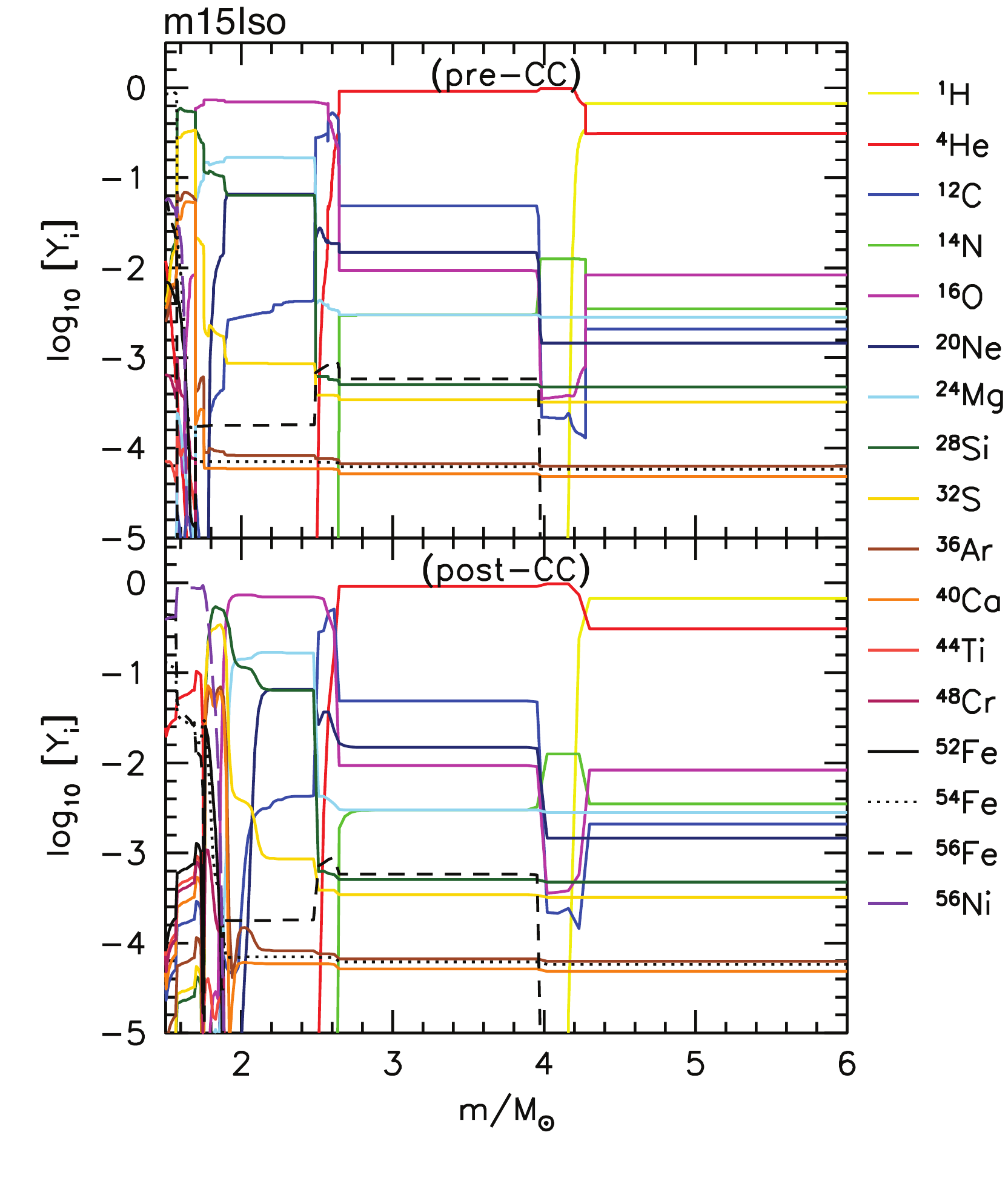}
\includegraphics[width=0.5\textwidth]{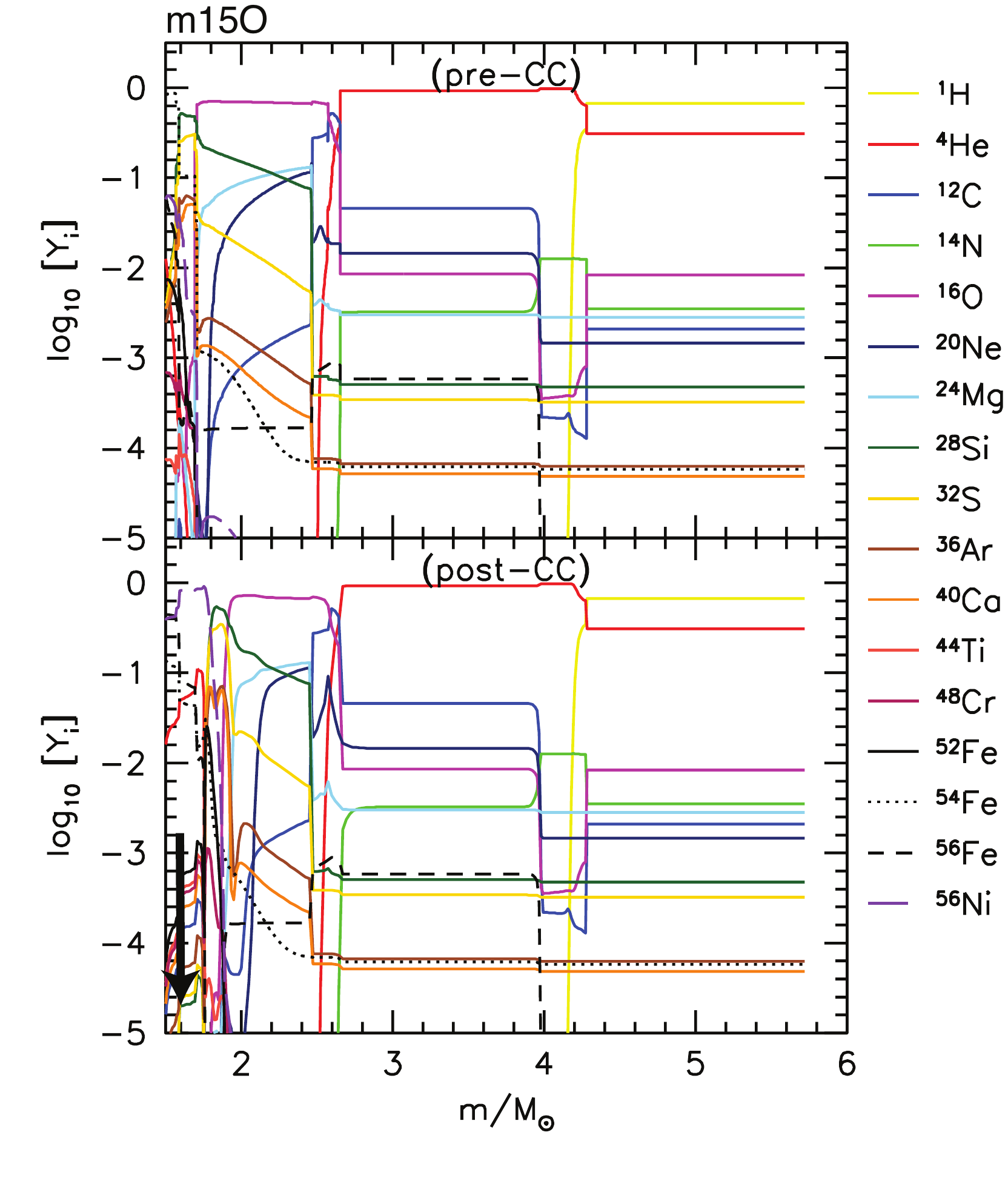}
\caption{{\it Left}: Composition profile for the 15M$_{\sun}$ model with
steady mass loss ({\tt m15Iso}). 
The upper panel shows the compostion of the inner 
6M$_{\sun}$ prior to core collapse, while the lower panel shows the 
composition for the same elements after core collapse. There is no mixing
assumed in these models. {\it Right}: Same as in the left-hand panel,
but for the model that underwent extreme mass loss during core oxygen
burning ({\tt m15O}). After evolving the models in the lower panels for 
400 years with our
{\tt ChN} code, $\sim$ 1M$_{\sun}$ of ejecta are shocked in the isotropic
wind model, while nearly all the ejecta in the highly stripped model
are shocked, as indicated by the arrow in the bottom righthand panel.}
\label{fig:comp}
\end{minipage}
\end{figure*}

For each presupernova model, we assume an explosion energy of 10$^{51}$ erg.
We choose to excise the inner 1.5M$_{\sun}$ (the approximate mass of the iron
core; Table~\ref{tab:mesa}) from the progenitor model, 
and spread the energy across 0.1M$_{\sun}$ of ejecta, corresponding
to a radial distance of $\approx$ 200 km above the proto-neutron star. 
The energy deposition
lasts for 100 ms. In multidimensional studies, typical core bounce
timescales are 100--200 ms \citep[c.f.,][]{ott08}, 
so our choice of 100 ms is appropriate. 
The final composition and structure of the
ejecta are both sensitive to the explosion energetics, but we chose 
values that are typical for core collapse supernovae, and consistent
with 1D models of this type \citep{morozova15}. The explosion is
followed to an age of 100 days. 

We plot in Figure~\ref{fig:comp}
the initial and final compositions for two of the models ({\tt m15Iso} 
and {\tt m15O}). As seen in these plots, the composition of the models
both pre and post core collapse are similiar. In both models, and
in model {\tt m15C} which is not shown, the chemical compositions of the
He-rich shell, located between mass coordinates of approximately 2.6 and
4.2M$_{\sun}$ are virtually identical, as are the compositions of
the H-rich shell, exterior of 4.2M$_{\sun}$. The differences in the
models are how far the H-rich shell extends. Interior of the He-rich
shell, differences in composition exist in the O-rich shell between 
1.8 and 2.6M$_{\sun}$, but these differences are likely due to 
differences that arise during the stellar evolution. Interior to this,
the compositions appear identical. It is not entirely surprising that
the compositions are so similar, since the same explosion conditions 
are applied to all three models. It is worth noting that the differences
in the EOS seen in Figure~\ref{fig:eos} correspond with locations
of shell boundaries. The difference around 4.2M$_{\sun}$ corresponds to
the boundary between the He- and H-rich layers, and the difference
around 1.8M$_{\sun}$ between the O-rich and Si-rich layers. Differences in
the EOS calculation appear insensitive to the boundary between the
O-rich and He-rich layers.

We do not report explicitly the final composition in our models 
(see section~\ref{sec:unc}). However, each model produces 
$\lesssim$ 0.2M$_{\sun}$ of $^{56}$Ni, and 2$\times$10$^{-4}$M$_{\sun}$
of $^{44}$Ti. Additionally, we estimate $\sim$ 0.5--0.8 M$_\sun$ of silicon, 
and 2--3 M$_{\sun}$ of oxygen are in the final ejecta models.

\subsection{Circumstellar Models}
\label{sec:csm}

In general terms, the circumstellar environment is dictated by the 
mass loss of
the progenitor, and the wind velocity of the lost material. Neither
parameter is completely constrained by observation \citep[for 
a recent review, see][]{smith14a}. Velocities
can vary from as little as 10 km s$^{-1}$ in a RSG, to $\gtrsim$
1000 km s$^{-1}$ in a Wolf-Rayet star. Mass loss rates can vary from
$\sim$ 0.1-10 M$_{\sun}$ yr$^{-1}$ in a Luminous Blue Variable eruption, 
to as low as 10$^{-7}$ M$_{\sun}$ yr$^{-1}$ in a Helium 
star\footnote[6]{Velocities and mass loss rates are taken from Table~1 
of \citet{smith14a}}, and binary interactions and rotation can 
act to further enhance the mass loss. 

As discussed in Section~\ref{sec:mesa}, we assume three mass loss
scenarios. (1): a steady wind
up to core collapse; (2): the onset of a fast wind with substantial
mass loss during core carbon burning. This wind expands into the slower
red supergiant wind that evolves in the CSM during hydrogen and helium
burning, and persists for $\sim$ 5000 years, and the star loses a few
solar masses of material; 
(3): extreme mass loss during core neon and oxygen burning. This phase
lasts for $\sim$ 500 years; the star loses $\sim$ 6 M$_{\sun}$ of material
during this phase. 

{\tt MESA} does report the mass loss as a function of time, and the
mass loss is observed to vary with each timestep. This is entirely
expected, as it is a derived quantity from other stellar parameters
which are also functions of time. However,
the dominant timescale in the CSM is the cooling time of swept-up shocked
CSM, $\sim$ 20 years. This can be much longer than the relevant timescale in
{\tt MESA}, which is driven by the core burning and is $\sim$ a few years
during carbon burning, and of order seconds during oxygen burning. 
In light of this, we adopt a hybrid approach.

We model the CSM as a power law wind which is formed by the progenitor
during H- and He-burning. This forms the CSM into which we evolve the
other models, and to which we compare the other models. For the
isotropic wind model, we adopt an average 
$\dot{M}$ $\approx$ 5$\times$10$^{-6}$
M$_{\sun}$ yr$^{-1}$, with a wind speed of 15 km s$^{-1}$. This is the
average value derived from {\tt MESA} output, over the life of the star. 
This wind will
form a shell of cooled, swept up ISM, but we estimate that for an initial ISM
density of 0.3 cm$^{-1}$, the shell radius is $\gtrsim$ 3.5 pc for the 1.5 Myr
evolution of the main sequence and helium burning phases, 
well beyond where the 
SNR shock will be after the 400 year post explosion evolution 
explored in this paper. 

We model the enhanced and extreme mass loss cases in similar manners.
We evolve these two cases into the steady wind produced during the 
earlier phases of 
evolution. Model {\tt m15C} is evolved with a mass loss rate of 
10$^{-4}$ M$_{\sun}$
yr$^{-1}$ and a velocity of 1000 km s$^{-1}$ until the core carbon abundance is depleted
below 10$^{-3}$. Model {\tt m15O} is evolved with a mass loss rate of 
0.1 M$_{\sun}$ yr$^{-1}$ and a velocity of order
the progenitor escape velocity, $\sim$ 100 km s$^{-1}$. For both cases,
we assume a rise time in the wind of 10 years. For model {\tt m15C}, we 
model the subsequent evolution after core carbon burning with a steady
wind with mass loss rates 10$^{-5}$ M$_{\sun}$ yr and a wind velocity
of 10 km s$^{-1}$. For model {\tt m15O}, the extreme mass loss persists
until the onset of core silicon burning, which is a short enough phase that
another mass loss model is not employed.

We use the numerical hydrodynamics code VH-1 \citep{blondin93} to model the
evolution of the wind. VH-1 is a multidimensional general purpose 
hydrodynamics code which also forms the basis of our cosmic-ray hydrodynamics
code \citep{ellison07}. For the purposes of modeling the CSM, we have
included a routine to follow radiative losses in the shocked,
swept up CSM, using both collisional and non-equilibrium ionization 
curves from \citet{sutherland93}. 

\begin{figure}
\includegraphics[width=0.5\textwidth]{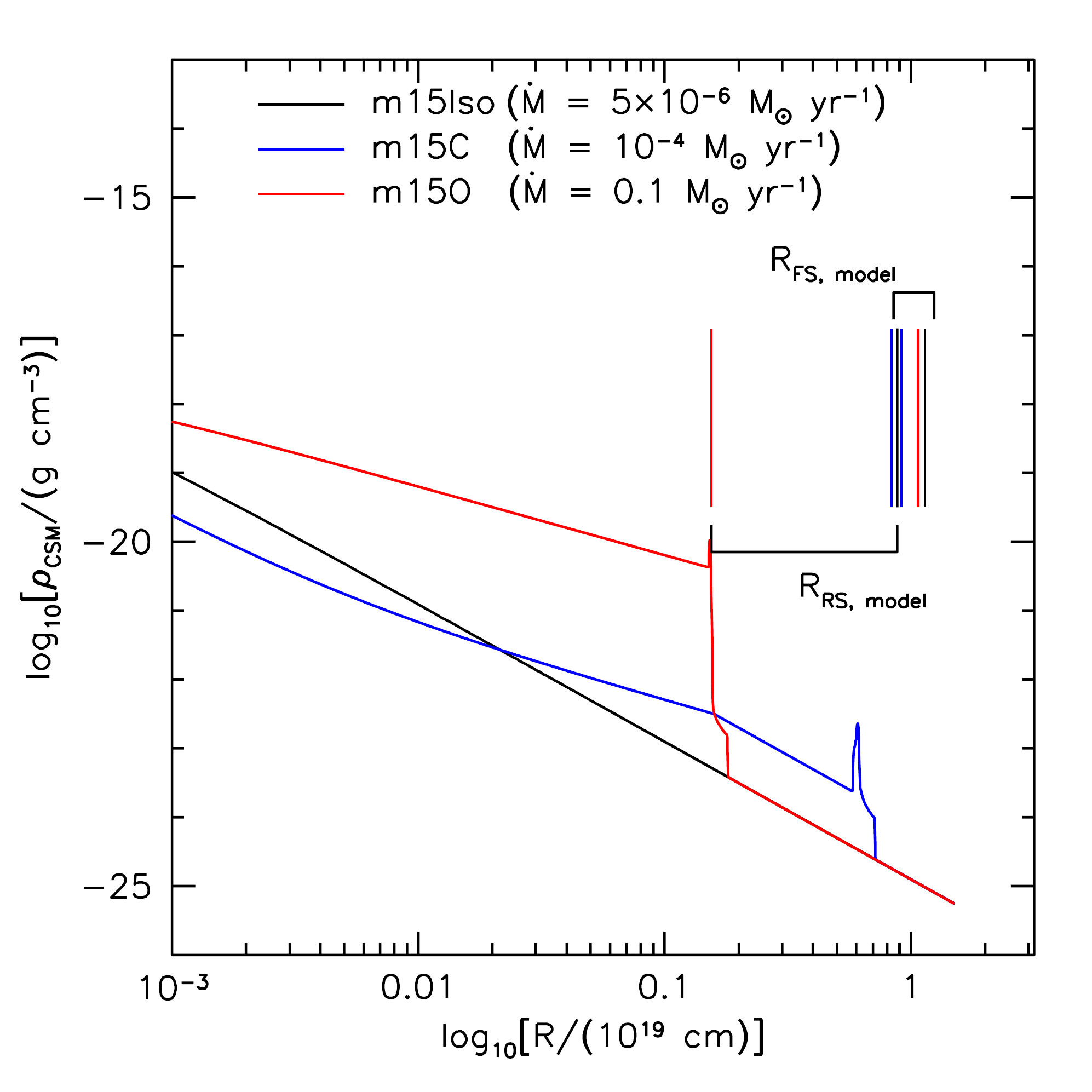}
\caption{Circumstellar environments for each progenitor model prior
to the remnant evolution. The black curve shows the CSM for model 
{\tt 15mISO}, while the blue curve corresponds to model {\tt 15mC}
and the red curve to model {\tt 15mO}. To accentuate the differences
in the structure of the environment, we plot the base-10 logarithm
of the radius. The final position of the SNR forward shock for each
model is labeled. The curvature in model {\tt 15mC} is a result in
a ramp down in the mass loss rate $\approx$ 1000 years before the
end of the simulation.}
\label{fig:wind}
\end{figure}

The density profiles of the modeled circumstellar environments 
are shown in Figure~\ref{fig:wind}. 
While the isotropic case follows the standard
$\rho_{\mathrm{CSM}}$ $\propto$ r$^{-2}$, a radiatively cooled CSM shell 
forms in the case of the 10$^{-4}$ M$_{\sun}$ yr$^{-1}$ wind. 
However, in the extreme mass loss
case, the shell does not cool radiatively before the simulation ends.
In this case, the progenitor will explode within a few days of the
exhaustion of the oxygen core. For reference, we mark the positions of the
forward and reverse shocks at t$_{\mathrm SNR}$ for each model of 
SNR evolution, discussed in the next section.

\subsection{Remnant Evolution Models}
\label{sec:chn}

 Lastly, we model the evolution of the ejecta discussed in 
 Section~\ref{sec:snec} into circumstellar profiles discussed in 
 Section~\ref{sec:csm}. We use our cosmic ray hydrodynamics code, 
 hereafter called {\tt ChN} to model the evolution of the ejecta
 to an age of $t_{\mathrm{SNR}}$ = 400 yr. {\tt ChN} is a Lagrangian
 hydrodynamics code that includes a prescription for diffusive shock
 acceleration \citep[DSA;][]{ellison07,lee12}. We have modified the code to 
 include the effects of DSA on non-equilibrium ionization 
 \citep{patnaude09,patnaude10} and have coupled the code to supernova
 ejecta models \citep{lee14,patnaude15}. We have also included radiative
 losses via forbidden line cooling \citep{lee15}. This effect will
 be important in the evolution of the SN shock with a nearby CSM shell, or
if we choose to model the radiative shock that could form in the ejecta
during early supernova evolution \citep{nymark06}. However, we begin
our simulations at an age of 5 years, and over  the lifetime of the 
simulation the shocks remain adiabatic, so we do 
 not consider the radiative shock model presented in our previous 
work here. Since {\tt ChN} couples 
 nonlinear particle acceleration to the SNR shock dynamics, we are able
 to reproduce the broadband thermal and nonthermal emission \citep{ellison10,
 ellison12,castro12,slane14,lee13}. The diffusive shock acceleration process
is an integral part of {\tt ChN}, and some injection of thermal particles
into the acceleration process is always assumed. Here we set the injection
parameter to the test particle limit, though we note that the
interaction of a strong shock with a massive CSM shell or cloud will lead
to enhanced particle acceleration \citep[e.g.,][]{ellison12,lee14}, and the
 differing CSM configurations, combined with the differing ejecta profiles
 and compositions, may result in differences in the broadband nonthermal
 emission. The study of nonthermal emission in evolving supernovae is
sufficiently broad that we defer its study to future papers.

\begin{figure*}
\begin{minipage}[c]{1.\textwidth}
\includegraphics[width=0.5\textwidth,viewport=205 20 548 548]{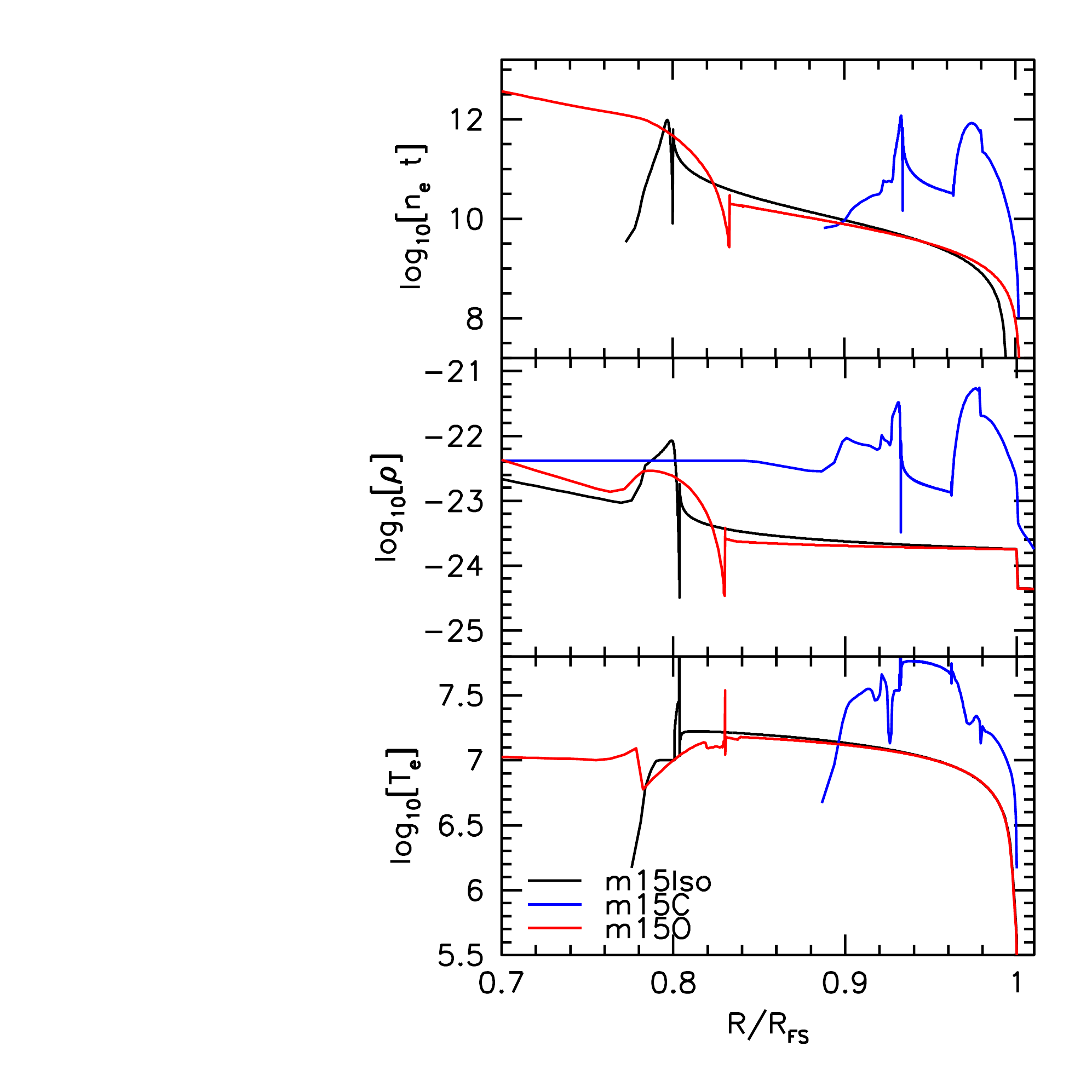}
\includegraphics[width=0.5\textwidth,viewport=205 20 548 548]{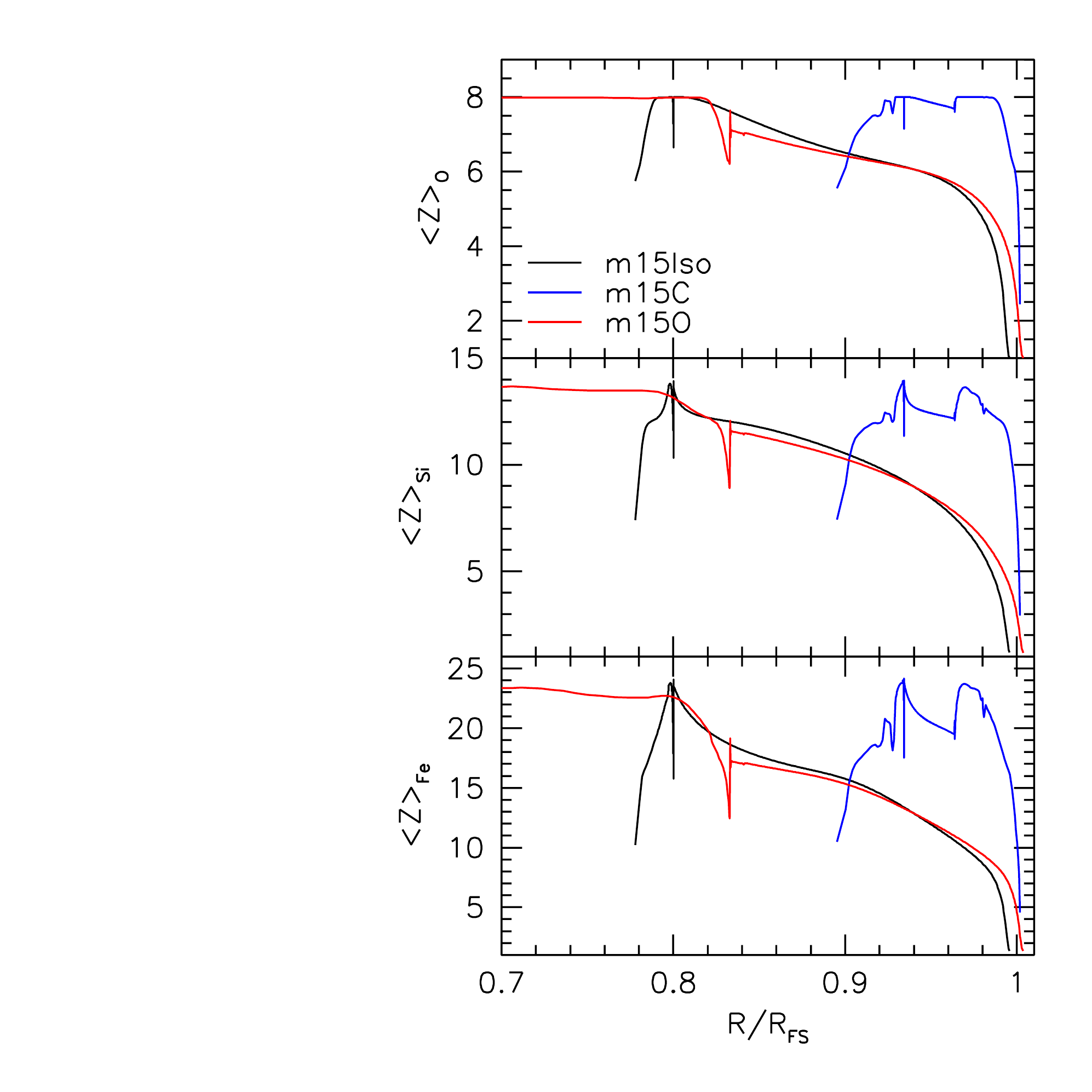}
\caption{{\it Left}: Hydrodynamical state of each model after 400 years
of evolution. The top and bottom panels show the ionization age ($n_e t$) 
and electron temperature ($T_e$) of shocked material only, while the middle
panel shows the density for shocked and unshocked material. The temperature
spikes seen in the plot of $T_e$ are due to contact discontinuities in the
1D model. These spikes coincide with regions of low density and thus do
not contribute to the overall emission. {\it Right}: Average charge state
for oxygen (top), silicon (middle), and iron (bottom).}
\label{fig:crsims}
\end{minipage}
\end{figure*}

 We simulate the SNR shock evolution to an age of 400 years. Using the
 time-dependent ionization balance, we compute the thermal X-ray 
 emission from the shocked CSM and ejecta. {\tt ChN} has the capability
 to compute spectra from {\tt APED} \citep{foster12}, or from a more primitive 
 X-ray emission code discussed in \citet{patnaude10}. While our
previous studies have made use of the code discussed in \citet{patnaude10},
we feel that the more thorough treatment provided by {\tt APED} will
make our simulations accessible to 
 future high-resolution X-ray spectroscopy missions. In 
Figure~\ref{fig:crsims}, we plot the final profile of each model
after 400 years. We also plot the final average charge state for
oxygen, silicon, and iron. 

\begin{figure*}
\begin{minipage}[c]{1.\textwidth}
\includegraphics[width=\textwidth]{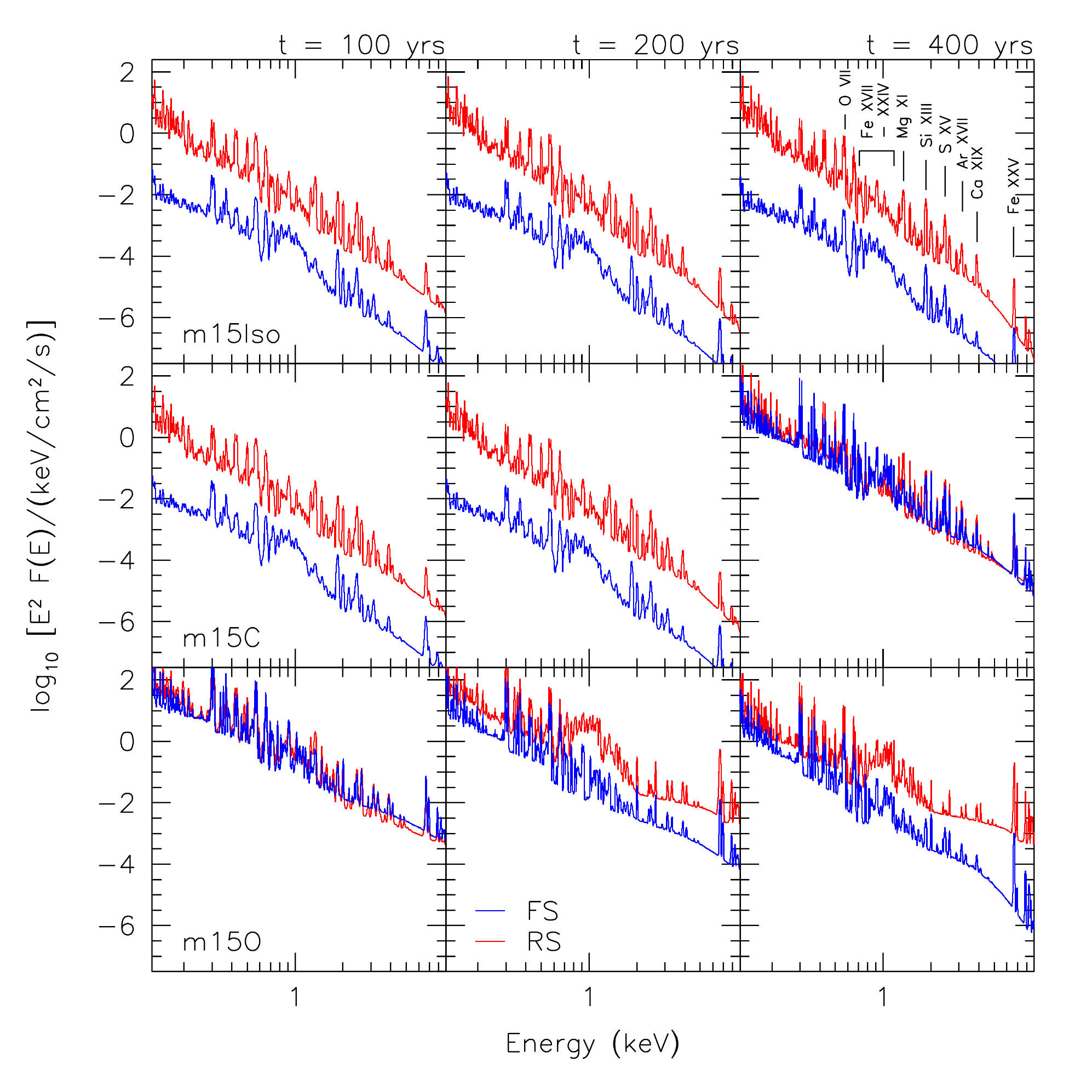}
\caption{Simulated X-ray spectra from the forward and reverse shock
for models {\tt m15Iso}, {\tt m15C}, and {\tt m15O}, at ages of 100, 200, 
and 400 years. Bright He-like lines and lines from \ion{Fe}{17}--\ion{Fe}{24}
are marked in the top right panel.}
\label{fig:xrayspec1}
\end{minipage}
\end{figure*}

\begin{figure*}
\begin{minipage}[c]{1.\textwidth}
\includegraphics[width=\textwidth]{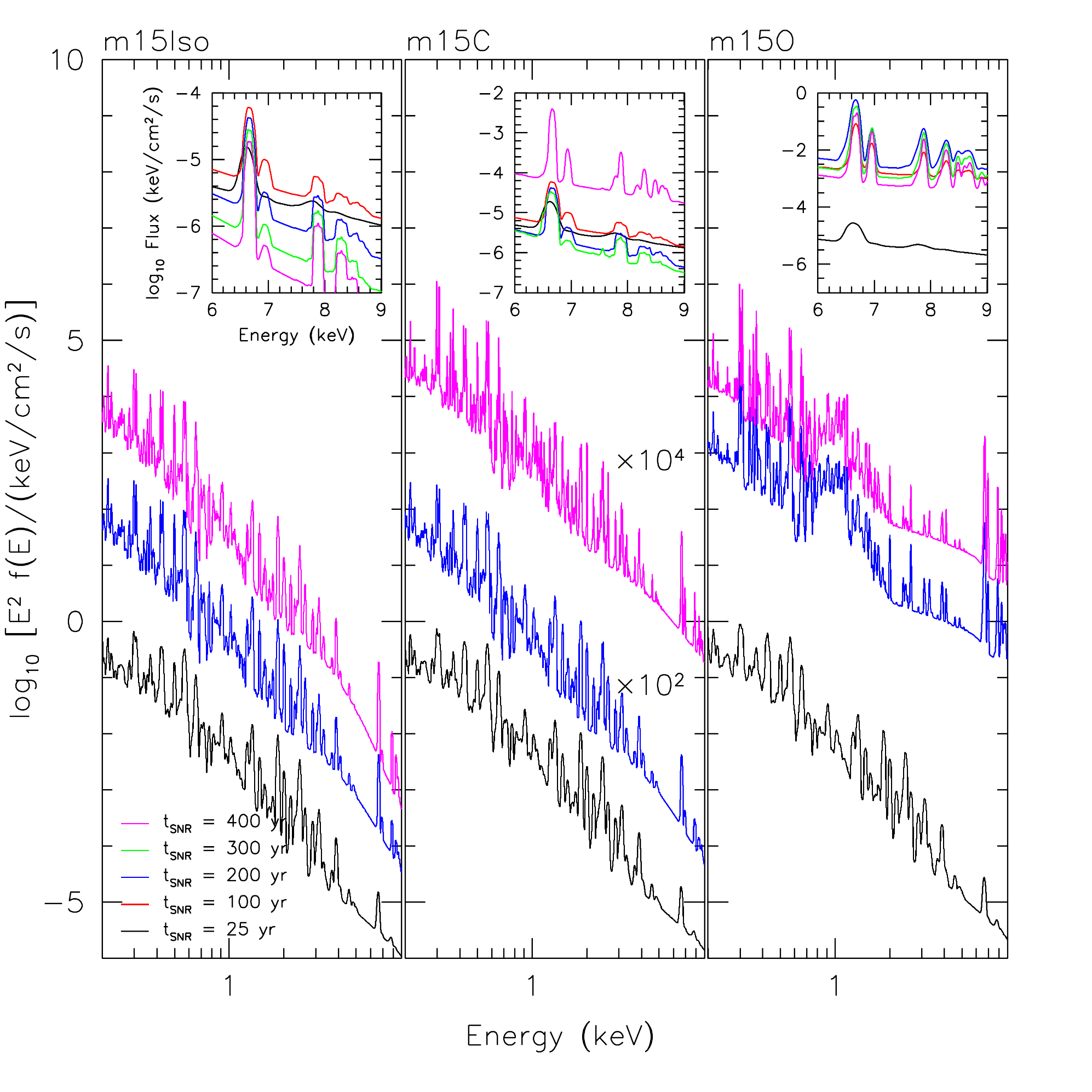}
\caption{Integrated X-ray spectra for each model at ages between 
25 and 400 years. In the main panels we plot the spectra at ages of 
25, 200, and 400 years. The spectra at 200 and 400 years are scaled
relative to the spectra at 25 years. The evolution of emission around 
Fe-K is shown in the insets, and represent absolute fluxes. We plot
the evolution of the Fe-K emission at 25, 100, 200, 300, and 400 years.}
\label{fig:xrayspec2}
\end{minipage}
\end{figure*}

We plot the evolution of the synthetic X-ray spectra
 from each model in Figures~\ref{fig:xrayspec1}--~\ref{fig:xrayspec2}. 
We have assumed an SNR distance of
 1 kpc. The spectra have been both thermal and Doppler broadened \citep{lee14},
and for clarity, we do not include the effects of interstellar absorption,
which can be significant below 1 keV.

\subsection{Model Uncertainties}
\label{sec:unc}

While we are not aiming to model any particular remnant in detail,
each component of the model chain has uncertainty associated with both
the input data and the chosen physics. Statistical uncertainty in
measured nuclear and atomic cross sections are discussed in detail
in their original source papers, referred to in the preceeding subsections. 
Here we aim to understand the uncertainty that is inherent in our
choice of input physics and parameters. We qualitatively
summarize the uncertainty below.

For our stellar evolution models, we do not account for rotation or
magnetic fields, and we choose a limited nuclear reaction network. The
choice of nuclear reaction network can impact the final abundances of
key elements such as oxygen and silicon in
the core at core collapse by as much as 30\% \citep{farmer16}. The exact
details of the mass loss mechanism remain poorly understood
\citep{smith14a}. The mass loss rates we choose, both the quasi-steady
rate in the isotropic model, as well the enhanced rates span 
a parameter space that is broadly consistent with observed rates for
steady and episodic mass loss, but given the one-dimensional 
nature of our models, do not account for effects such as clumping in the
wind. Additionally, we have smoothed the CSM density with a Gaussian kernel 
in an attempt to smear out the sharp transitions that occur around the
CSM shell boundary. In reality, the transition between the powerlaw
wind and the CSM shell may be more complicated.

In the explosion phase of our modeling, the choice of mass cut, thermal
bomb duration, and bomb spread can all affect the nucleosynthesis. 
\citet{young07} studied variations in nucleosynthetic yields in 
one-dimensional explosion models and came to the conclusion that high-$Z$
element production is sensitive to the explosion energy, and that yields
may 
differ by as much as 50\% between thermal bomb and piston driven explosions. 
In 
essence, the yields are non-unique for a given progenitor, and can
vary based on how the energy is deposited in the progenitor. The progenitor
masses at core collapse differ for our three models, though we choose
the same explosion energy, bomb spread, and heating duration. Since the
nucleosynthesis is sensitive to this and the core density and composition,
and since these do not vary across the three models by much, the
final abundances interior of the helium core are very similar, with
each producing $\lesssim$ 0.2 M$_{\sun}$ of $^{56}$Ni. A more dynamic reaction
network would probably lower this nickel mass by quite a bit \citep{young07},
though 0.2$M_{\sun}$ of $^{56}$Ni is consistent with yields expected
from some core collapse supernova models \citep{young07}.
For the remnant modeling, we do not consider non-linear shock acceleration
effects, which can alter the dynamics and emitted spectrum 
\citep{ellison07,patnaude11,lee12}. 

Finally, we comment on the limitations of our one-dimensional modeling. As 
discussed above, we don't include the effects of rotation in the stellar
evolution modeling. This can alter the mixing between layers and the 
treatment of convection during the stellar evolution. In the CSM, 1D 
models result in CSM shells, instead of a web of tenuous wind peppered
by condensed clumps. For the explosion and remnant evolution modeling,
1D models are unable to follow the effects of mixing due to the 
Rayleigh-Taylor (RT) or Kelvin-Helmholtz instabilities. \citet{duffell16}
recently presented a prescription for the 1D RT instability, and
we will incorporate these effects in subsequent studies. We have also
not included the bulk mixing of ejecta via convective instabilities
during the explosion. This will alter the abundances in the outer
layers of the star, resulting in changes to the emitted X-ray spectrum
from the ejecta. 

Our models represent a first attempt to follow the
complete evolution of a massive star from the pre-main sequence
through the remnant phase. Each phase of evolution takes as input 
parameters derived from a prior stage, allowing for a quasi self-consistent
study of how massive star evolution affects the remnants we observe
today. As is clear from the uncertainties discussed in this 
section, an end-to-end supernova simulation requires a number of 
approximations and assumptions. Nevertheless, we show below that
meaningful constraints on the ``hidden'' supernova properties can
be deduced from SNR observations made centuries after the explosion.

\section{Modeling Results and Discussion}
\label{sec:results}

The principal output from {\tt ChN} includes the blastwave dynamics,
as well as the broadband thermal and nonthermal emission. The results of
our simulations are summarized in Table~\ref{tab:results}. For each 
model, we list
the swept up mass and blastwave radius, the amount of shocked ejecta,
and the bulk energy centroid for the He-like state of iron. We discuss
the dynamical and spectral results below. 

\subsection{Model Differences}

As seen in Table~\ref{tab:results}, after 400 years, the blastwave radii
for models {\tt m15Iso} and {\tt m15O} are virtually identical, despite
different stellar evolutionary histories. On the otherhand, model {\tt m15C}
is $\sim$ 15\% smaller over the same time period. The positions of the 
forward shock, relative to the CSM are shown in Figure~\ref{fig:wind}. 

The blastwave radii are similar for {\tt m15Iso} and {\tt m15O}, 
but the amount
of swept up mass and shocked ejecta are remarkably different. {\tt m15Iso}
has swept up less than a solar mass of material in 400 years, and only shocked
$\sim$ 2 M$_{\sun}$ of ejecta in that time. In contrast, in {\tt m15O} 
the blastwave
has shocked $\sim$ 8M$_{\sun}$ of CSM material, and has progressed all the
way into the center of the ejecta. Model {\tt m15C} represents an intermediate 
case, in that it has shocked about 2 and 6M$_{\sun}$ of 
CSM and ejecta material, 
respectively. Spectra from shock heated CSM and ejecta for each model are 
shown in Figure~\ref{fig:xrayspec1}, for ages of 100, 200, and 400 years, 
and differences in the amount of shocked material are readily apparent.

 As seen in Figure~\ref{fig:xrayspec1}, the spectral evolution for
models {\tt m15Iso} and {\tt m15C} are virtually identical over the first 200 years
of their evolution. At an age of 400 years, however, differences in their evolution 
become apparent, as the forward shock in model {\tt m15C} interacts with the
CSM shell (at $\approx$ $t_{\mathrm SNR}$ = 230 yr), and emission from shocked 
ejecta becomes comparable to that of shocked CSM. In contrast, at 100 years, the
emission from shocked ejecta and shocked CSM are comparable in model {\tt m15O}. 
In this
model, the shock interacts with the CSM shell at $\approx$ 40 years, and breaks
out $\approx$ 100 years later. By 200 years, the forward shock is well  
into the lower density pre-shell wind, so the emission from swept up CSM begins to 
drop, due to adiabatic expansion.

As listed in Table~\ref{tab:results}, both models {\tt m15C} and {\tt m15O} have swept
over a comparable amount of ejecta. However, the composition of the shocked ejecta is
quite different. At the time of core collapse, model {\tt m15C} still has a H-rich
envelope of mass $\sim$ 8M$_{\sun}$. The H-rich envelope for model {\tt m15O} is only
$\approx$ 2M$_{\sun}$. Additionally, as seen in Figure~\ref{fig:star}, exterior of 
R $\approx$ 10R$_{\sun}$, the density of model {\tt m15O} is lower than that of
either {\tt m15C} or {\tt m15Iso}. The progenitor of {\tt m15O} is not only more
compact than the other models, but it also has a lower density envelope -- the reverse shock
can travel all the way into the center of the ejecta after only 400 years. At 
400 years, model {\tt m15O} has already transitioned to the Sedov phase. This rapid
transition is likely aided by the dense CSM shell.

\begin{figure}
\includegraphics[viewport=190 10 550 550,width=0.5\textwidth]{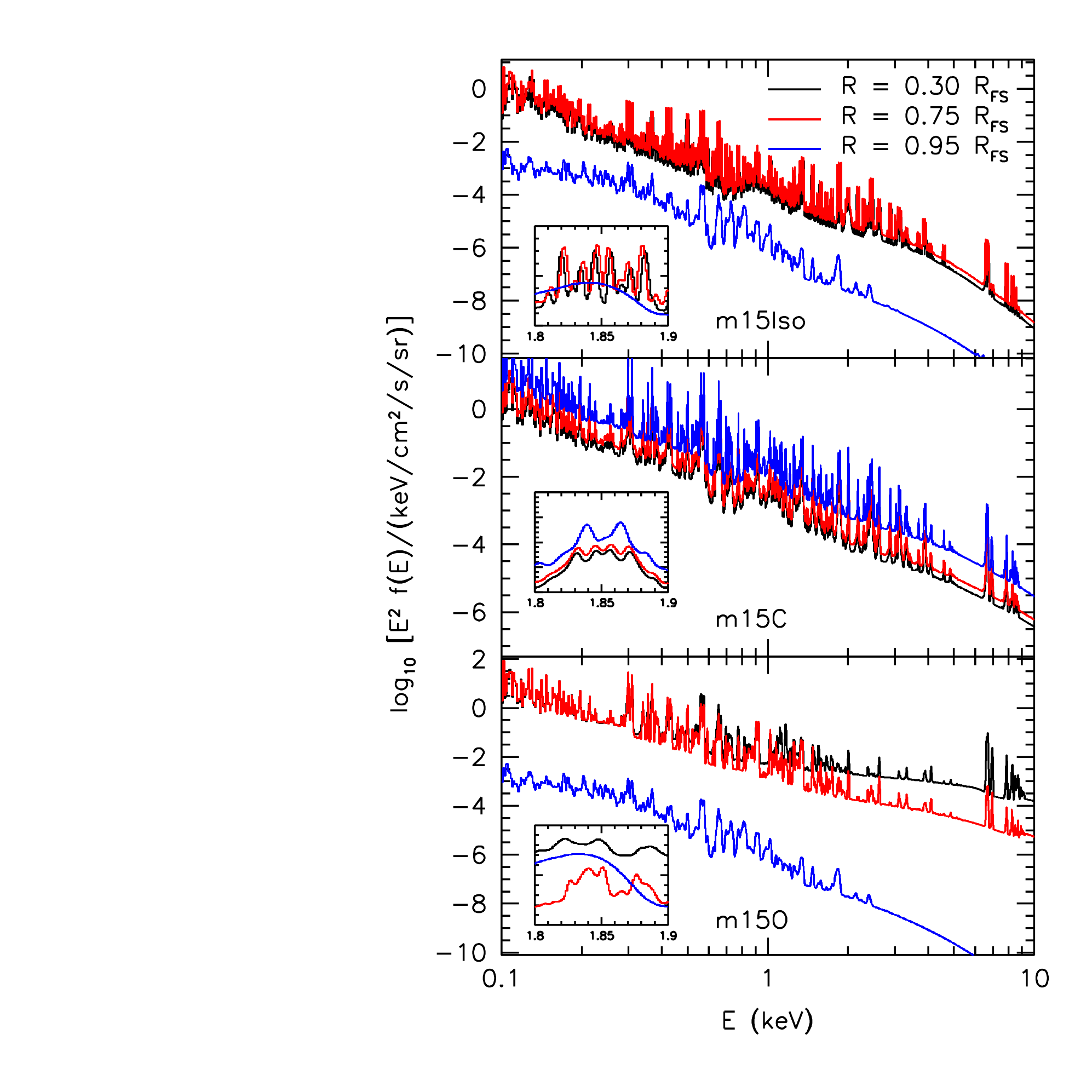}
\caption{Integrated spectra for models {\tt m15Iso} (top), {\tt m15C} (middle) and
{\tt m15O} (bottom), for projected radii of 0.3, 0.75, and 0.95R$_{\mathrm{FS}}$. 
Each radial extraction is of width $d\mathrm{R}/\mathrm{R_{FS}}$ = 0.1. The inset
regions show the line emission centered around \ion{Si}{13}. For ease of 
comparison, the line profiles have been scaled by an aribitrary amount. The Doppler
shifting of the line emission is seen most readily in models {\tt m15Iso} and {\tt m15C}.
The red-shifted lines have not been corrected for absorption.}
\label{fig:losspec}
\end{figure}

In Figure~\ref{fig:xrayspec2} we plot the time evolution
of the total thermal X-ray spectrum at 
$t_{\mathrm{SNR}}$ = 25, 200, and 400 years. 
For each model, we offset the spectrum from each epoch, 
for ease of comparison. We also show, inset, the evolution of the spectrum
from 6-7 keV, without an offset. Even with the offset in the y-axis,
the changes in the spectrum as the SNR evolves are apparent between the three models. In 
model {\tt m15O}, a sharp increase in Fe-L shell (\ion{Fe}{17} -- \ion{Fe}{24}) emission around 1 keV is
seen after 200 years, while by 400 years, much of the Fe emission comes from
K-shell emission. In contrast, as expected, there is a dramatic rise in the
total X-ray emission in model {\tt m15O} between 25 and 200 years. Even after the shock
breaks out of the shell, Fe-L shell and K-shell emission continue to increase over the
remainder of the SNR's evolution, probably as the reverse shock probes the deeper layers of the
ejecta. This is best exemplified when contrasting the bottom row of Figure~\ref{fig:xrayspec1} 
with the righthand column of Figure~\ref{fig:xrayspec2}: the X-ray emission from reverse shock 
heated material (red curves of Fig.~\ref{fig:xrayspec1}) rises
between 100 and 200 years, and only declines a bit over the next 200 years. In contrast, emission
from shocked CSM drops steadily across the three epochs in these two models. 
The late time Fe emission shown in Figure~\ref{fig:xrayspec2} from
model {\tt m15O} arises predominantly from shocked ejecta.

We plot the absolute line fluxes for K-shell emission in the inset panels of Figure~\ref{fig:xrayspec2},
and list the centroid energies at $t_{\mathrm{SNR}}$ = 400 years, in Table~\ref{tab:results}. Model
{\tt m15O} results in a considerably higher line centroid at 400 years, than the other two
models ($\sim$ 10 eV greater). As seen in the inset, the absolute line fluxes are nearly 2 orders
of magnitude higher as well. This is expected -- the blastwave has both swept up more material in
the CSM {\bf and} shocked more ejecta than the other models. The overall emission measure for 
model {\tt m15O} is higher, producing a
higher overall flux, and the ionization timescale $\tau = \int n_e(t) dt$
is much larger as well, resulting in a higher ionization state (see Figure~\ref{fig:crsims}). 

\begin{figure}
\includegraphics[viewport=190 10 550 560,width=0.5\textwidth]{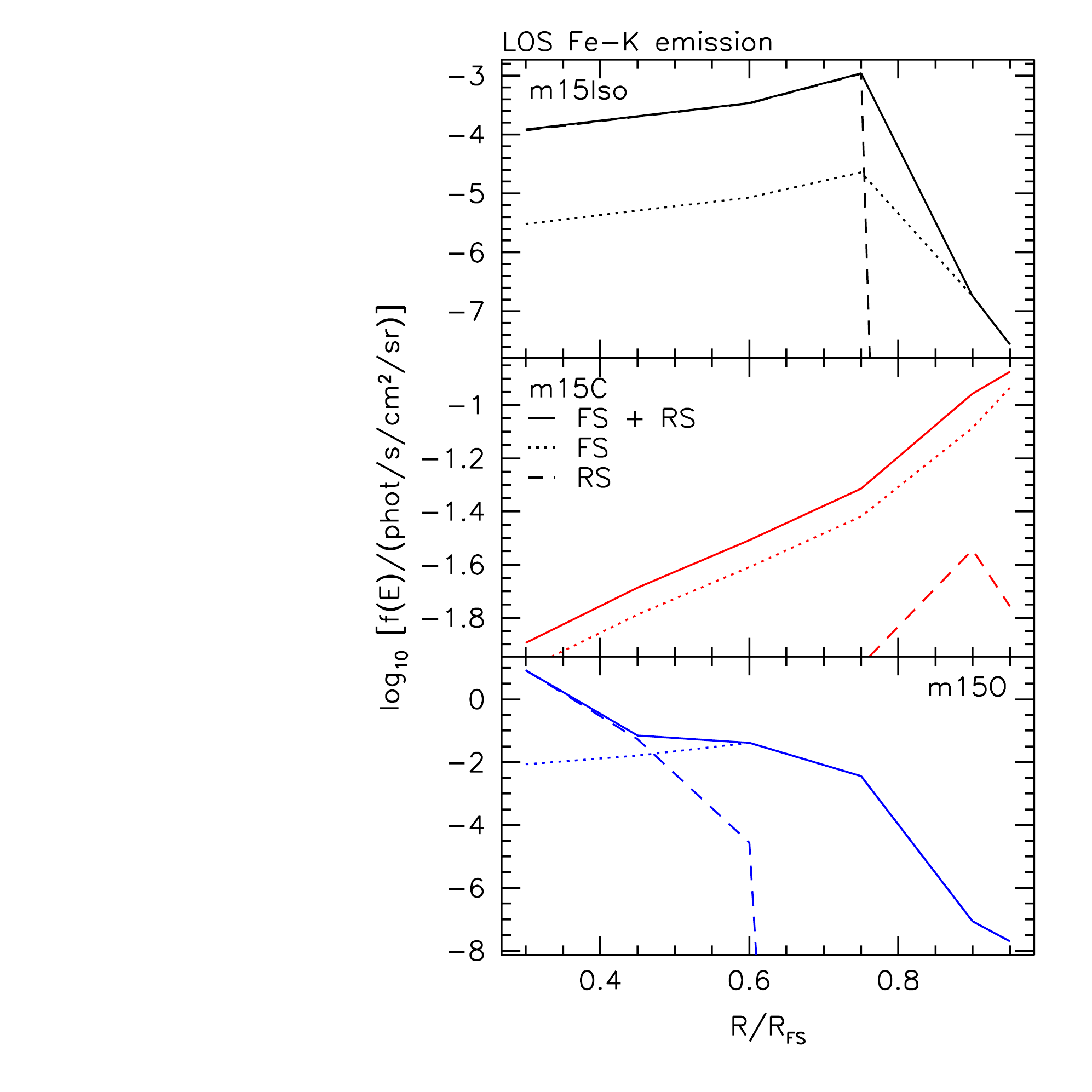}
\caption{Line of sight projected emission from 6.4-6.8 keV, highlighting emission from Helium-like
iron. The top panel corresponds to model {\tt m15Iso}, the middle to {\tt m15C}, and the bottom
to {\tt m15O}. Each panel shows the contribution to the total flux from both the forward and reverse
shock. In the case of model {\tt m15C}, the shocked ejecta have caught up to the shocked CSM, as the 
forward shock moves through the CSM shell, over a radial distance that is unresolved by the choice
of radial binning. All the plots have been normalized to the forward shock radii for each model.}
\label{fig:losfek}
\end{figure}

Finally, in Figures~\ref{fig:losspec} and~\ref{fig:losfek}, we plot the line of sight
integrated spectra for three fiducial radii, and the projected 6.4-6.8
keV emissivity. We include both Doppler and thermal broadening in 
the spectral computations. At an age of 400 years, Doppler shifts are still detected towards the
center of the SNR, allowing for discrimination between blue-shifted and red-shifted ejecta out
to radii of 0.75R$_{\mathrm{FS}}$. We highlight emission around \ion{Si}{13}. As seen in 
Figure~\ref{fig:losspec}, there does not appear to be any emission from this state of
silicon in model {\tt m15Iso} and {\tt m15O} at a radius of 0.95R$_{\mathrm{FS}}$. This is
confirmed by the low charge state of silicon near the forward shock in these models, seen 
in the middle panel of Figure~\ref{fig:crsims} (right), and in contrast to the much higher
average charge state of silicon in model {\tt m15C}.

For the line of sight Fe-K emission, models {\tt m15Iso} and {\tt m15O} show expected profiles. Interior
to the contact interface, the Fe-K emission is dominated by emission from the reverse shock. Exterior
to this, emission is from the shocked CSM only. In the case of model {\tt m15C}, emission
is primarily from forward shocked material everywhere. This is likely due to the strong interaction
between the blastwave and the CSM shell. Interestingly, the reverse shock heated material is projected
nearly to the forward shock. This is due to the fact that the forward shock is strongly decelerated
in the shell. The radius of the contact interface in our model is 0.93R$_{\mathrm{FS}}$. For 
Figure~\ref{fig:losfek}, we choose extraction regions with resolution 0.1R$_{\mathrm{FS}}$. While
the contact interface is very close to the forward shock, our choice of extraction binsize results
in the outermost bin of shocked ejecta being projected to the radius of the forward shock.

\subsection{Implications for Progenitor Identification}
\label{sec:id}

In the absence of a light echo spectrum which can be compared to 
template spectra for core collapse supernovae, relating a remnant
back to its progenitor evolution remains tricky. As already discussed
here, important mass loss processes can be triggered by several
channels, including binary interaction and enhanced or episodic
mass loss. Additionally, the once clear roadmap between progenitor and 
supernova type is more muddled, as supernovae are now observed
to migrate between types as they evolve \citep[e.g.,][]{milisavljevic15}.

In specific terms, there have been several attempts to connect 
supernova remnants back to their progenitors Most recently, \citet{katsuda15}
detected thermal X-ray emission from the synchrotron dominated 
SNR RX J1713.7--3946. They found that the measured abundances 
favored a low mass progenitor, and that it likely lost much of 
its mass through binary interaction. This is at odds with 
previous work which considered a massive O star that carved out a large
windblown bubble that the progenitor proceeded to explode into 
\citep[e.g.,][]{ellison12}.

More generally, \citet{chevalier05} looked at the  morphologies of 
several Galactic SNR, combined with qualities of their central 
compact objects, and typed several as IIPs, IILs, or IIbs, dependent
upon the amount of mass estimated to have been lost during stellar
evolution. 

\begin{figure}
\includegraphics[viewport=200 10 550 560,width=0.5\textwidth]{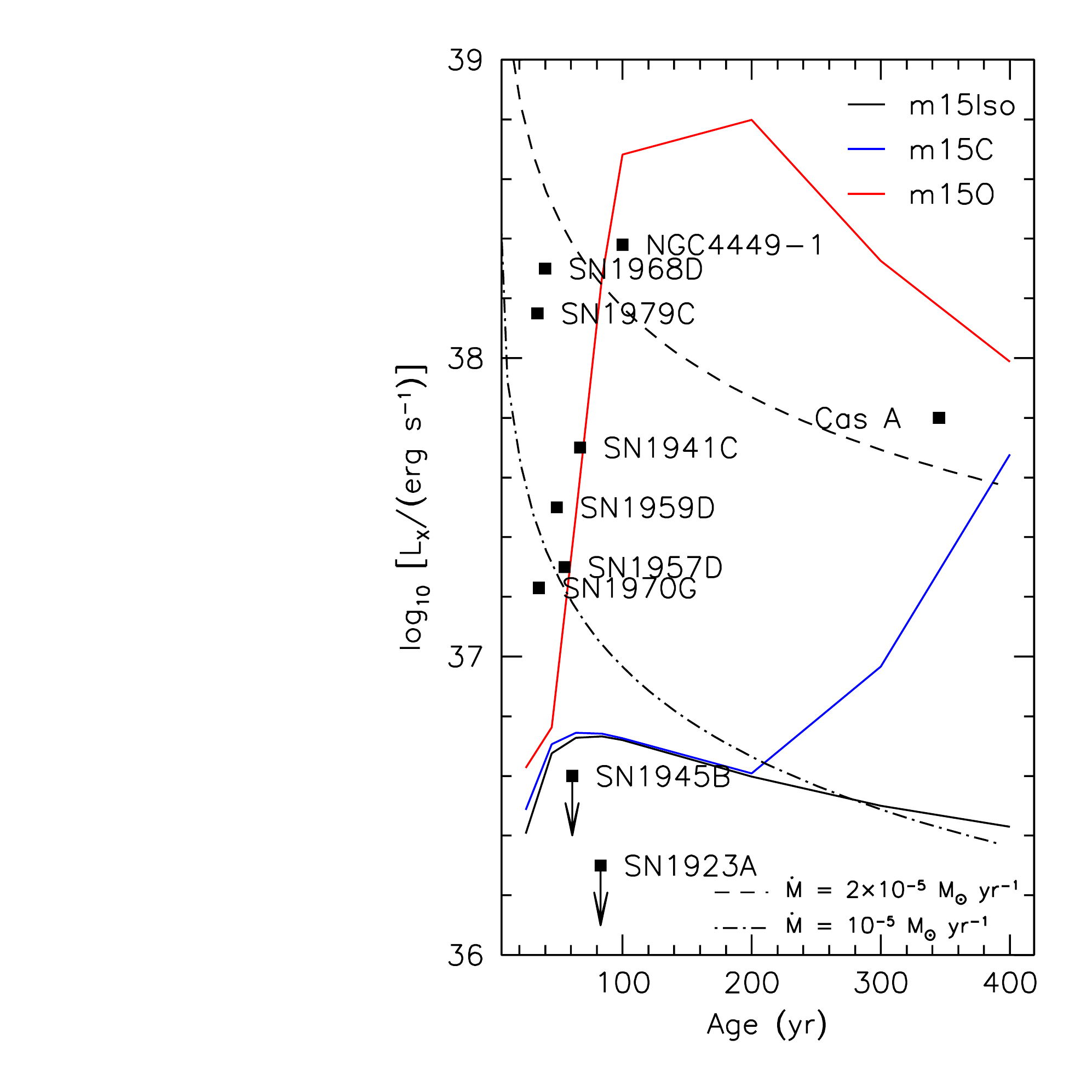}
\caption{0.5 - 10.0 keV X-ray luminosity for each model. Also shown are
the approximate X-ray luminosities for several historical supernova remnants,
as well as the expected X-ray luminosity due to free-free emission 
for mass loss rates of 10$^{-5}$ and
2$\times$10$^{-5}$ M$_{\sun}$ yr$^{-1}$, assuming isotropic mass loss
\citep{immler03}.
Data are taken from \citet{patnaude03,stockdale06,soria08,patnaude11,
long12,ross17}.}
\label{fig:ltc}
\end{figure}

 {\tt ChN} has the capability to compute X-ray lightcurves. 
 \citet{dwarkadas12} compiled the lightcurves for all known X-ray 
 supernovae, and showed that many deviate from the expected 
 $L_X$ $\propto$ $t^{-1}$ behavior. In Figure~\ref{fig:ltc}, 
we plot the 0.5--10 keV
 lightcurves for the three models. The lightcurves are distance 
 and absorption independent.

As seen in Figure~\ref{fig:ltc}, there is a steady decline in the
X-ray emission in model {\tt m15Iso}, while the models with non-steady
mass loss show different behavior - model {\tt m15O} shows a sharp
increase in luminosity due to the early interaction between the blastwave
and CSM shell, while model {\tt m15C} follows model {\tt m15Iso} 
closely before the shock--shell interaction. We have overlaid the current
X-ray luminosity for select historical core collapse supernovae, as 
well as theoretical curves for the X-ray emission as a function of time for
a range of mass loss rates and wind velocities \citep{immler03}. 
While we are not aiming to model any particular SNR, it is worth noting
that the data from any particular SNR are neither inconsistent with
the luminosity predictions from our models, nor the predictions from
self-similar models such as those shown in Figure~\ref{fig:ltc}. 

\citet{dwarkadas12} published historical lightcurves for 41 X-ray 
supernovae. They showed that the lightcurves do not decline as $t^{-1}$
as would be expected from isotropic mass loss. They argued that some of
this is due to how the X-ray temperature changes as the blastwave evolves,
resulting in differing {\it observed} emission as the peak of the emissivity function 
changes, but some of the trends they observe in the lightcurves may
also be due to the structure of the circumstellar environment. 
The theoretical curves presented in Fig.~\ref{fig:ltc}
represent contributions from shocked CSM only, and do not account for
the delayed rise in X-ray emission from shocked ejecta 
\citep[c.f., Figure~1 right of ][]{patnaude15}. 
Qualitatively, the luminosity in the
isotropic mass loss models are similar to our models, but they assume
a steady decline in the CSM density, which is clearly
not the case for our models {\tt m15C} and {\tt m15O}. Of additional
interest are the large differences in luminosity between observations
and our models (and the self similar models overplotted in 
Figure~\ref{fig:ltc}). As seen in \citet{dwarkadas12}, the X-ray luminosity
for several of their supernovae varies between 10$^{38}$ and 10$^{40}$
erg s$^{-1}$ over the first few decades of supernova evolution. These
observed luminosities are much higher than what we see in our models at 
early epochs. We
interpret this as meaning that the mass loss rates are $>$ 10$^{-5}$
M$_{\sun}$ yr$^{-1}$ prior to core collapse, or that dense CSM shells
exist at radii $<$ 10$^{17}$ cm. Observations of recent and not-so recent
SN support this 
\citep[e.g., SN~1996cr, SN~2005kd, and SN~2014C;][]{dwarkadas10,dwarkadas16,
margutti17}. Alternatively, the models presented here begin at an 
age of 5 years post core collapse. Choosing a starting age for our
simulation closer to the time of core collapse would likely result in 
higher X-ray luminosities
earlier in the evolution, in line with the self-similar predictions.

An interesting feature of Figure~\ref{fig:ltc} is the gap of observational
data between $\sim$ 50--100 year old extragalactic remnants, and the $\sim$
340 year old Cas A SNR. Remnants with ages of $\sim$ 100 years probe 
the mass loss history in the latter stages of the red supergiant
phase, an interesting
time in massive star evolution. Future X-ray observatories such as 
{\it Athena} will be able to access these epochs in young SNR such as 
SN~1957D and NGC~4449-1. 

When comparing the X-ray light curves in Figure~\ref{fig:ltc}, it becomes
clear that the integrated X-ray emission from a supernova or supernova
remnant does not tell the full story. At any one epoch, dissimilar mass 
loss rates can give similar $L_X$'s. Examining the dynamics and the 
detailed ionization state of the gas will break the degeneracy. For instance,
at ages of 400 years, model {\tt m15C} and the 
self similar model with a steady 2$\times$10$^{-5}$
M$_{\sun}$ yr$^{-1}$ differ in luminosity by less than 0.1 dex. At that
age, {\tt m15C} has a radius of 1.62 pc. In contrast, the blastwave radius for
a 400 yr old SNR with constant mass loss is $\sim$ 4 pc 
\citep[see Figure 1, right of][]{patnaude15}. 
From \citet{patnaude15}, the Fe-K luminosity for a model with mass loss
rate of 2$\times$10$^{-5}$ M$_{\sun}$ yr$^{-1}$ is estimated to be 
$\lesssim$ 10$^{42}$ photon s$^{-1}$. The luminosity in the Fe-K line
in model {\tt m15C} is estimated to be $\gtrsim$ 400$\times$ this. 
Even though the broadband X-ray luminosities between the models are
quite close, the details of the dynamics and ionization balance of the
shocked material tell a different story about how the blastwave interacted
with the pre-supernova environment.

\section{Conclusions}
\label{sec:conc}

We have presented the first quasi self-consistent 
models for the evolution of 15M$_{\sun}$ 
stars from the pre-main sequence through core collapse, and into
the remnant phase. To our knowledge, this is the first attempt to 
produce such an end-to-end simulation in a self-consistent 
fashion. We have followed the evolution of the remnant to an age
of 400 years, at which point one of our models, {\tt m15O}, has entered
the Sedov phase. The only difference between the three models is the 
mass loss history of the progenitor. We find that the mass loss in late
stages (during and after core carbon burning) can have a profound impact
on the dynamics and spectral evolution of the supernova remnant. While
our models are currently not tailored to any particular SN or SNR, we note
that:

\begin{itemize}

\item Extreme mass loss during core neon or oxygen burning can
result in CSM shells at distances $>$ 0.5 pc. While the shell is not 
in the immediate vicinity of the progenitor, the shock/shell interaction
will leave its imprint on the emitted X-ray spectrum centuries after the 
shock has broken through the shell.

\item Enhanced mass loss during post core helium burning phases can result in
CSM shells at radii of less than a couple of parsecs. These shells are created only a few
thousand years prior to core collapse. While we do expect them 
to collapse to thin shells due to radiative cooling, the shell will
persist through progenitor core collapse. Depending upon the energetics,
the blastwave will interact with the remnant of the shell up to a couple of hundred
years after core collapse, resulting in an increase in X-ray emission
from the shocked CSM. Our one-dimensional models, which do not follow
in detail the dynamical and radiative evolution of the shell after its
formation, which may lead to clumping or fragmentation of the shell, 
provide an upper limit on the amount of X-ray emission from
the shock-shell interaction in this scenario.

\item In \citet{patnaude15}, we postulated that enhanced mass loss 
in the years leading up to core collapse would result in increased X-ray
emission, with little impact on the late time dynamics. Our simulations
bear this out -- CSM shells close to the progenitor result in a sharp
increase in the X-ray emission up to a century after core collapse. However,
once the shock breaks through the shell, it accelerates and the forward
shock is dynamically similar to models with isotropic mass loss. We 
expect that this is due to the energetics of the explosion: in models
with enhanced or extreme mass loss, the specific energy of the ejecta
is higher than in the more massive model (by about a factor of 2). During
the early phases of the remnant evolution, the blastwave dynamics are not
strongly determined by the CSM structure, since the total mass in the
ejecta is half that of the isotropic mass loss model. This argues that when
considering the X-ray emission from supernova remnants, the mass loss 
history of the progenitor should be carefully considered. {\it Where} and
{\it when} the mass was deposited in the CSM can have a profound impact on
the evolution of the remnant.

\end{itemize}

Our models are not yet tailored to pinpoint the evolutionary history of any one
supernova remnant. However, given the high fidelity data that currently
exists for both evolved (t$_{\mathrm{SNR}}$ $\sim$ 1000 yr) and young 
(t$_{\mathrm{SNR}}$ $\lesssim$ 100 yr) SNRs, the progenitor mass loss
history can be reconstructed with reasonable precision 
\citep[see, e.g.,][]{dwarkadas12}. Thus, Galactic and extragalactic SNRs may 
now (or in the future) be probed as a class 
of objects. With the forthcoming advent of high spatial and spectral
resolution microcalorimeters on missions such as {\it Athena} and {\it Lynx},
we can probe the progenitor evolution of extragalactic SNe and SNR
by studying the detailed evolution of their spectra and comparing them
to our evolutionary models.

\acknowledgements

The authors wish to thank Frank Timmes, H.~Thomas Janka, Randall 
Smith, and Rob Fesen for fruitful discussions.
D.~J.~P. acknowledges support from the {\it Chandra} Theory Program
NASA/TM6-17003X, and NASA contract NAS8-03060. 

%%%%%%% references %%%%%%%%%

\begin{deluxetable}{lccccccr}
\tablecolumns{8}
\tablewidth{0pc}
\tablecaption{{\tt MESA} Initial and Final Model Parameters \label{tab:mesa}}
\tablehead{
\colhead{Model} & \colhead{M$_{\mathrm{Final}}$} & 
\colhead{M$_C$} & \colhead{M$_O$} & \colhead{M$_{Si}$} & \colhead{M$_{Fe}$} &
\colhead{R} & \colhead{$\dot{M}$\tablenotemark{a}} \\
\colhead{} & \multicolumn{5}{c}{M$_{\sun}$} & \colhead{$\log_{10}{R/R_{\sun}}$} &
\colhead{M$_{\sun}$ yr$^{-1}$}}
\startdata
m15Iso & 13.3 & 2.56 & 2.48 & 1.70 & 1.53 & 2.99 & 5 $\times$ 10$^{-6}$ \\
m15C   & 10.0 & 2.56 & 2.49 & 1.68 & 1.51 & 3.03 & 10$^{-4}$ \\
m15O   &  5.7 & 2.56 & 2.46 & 1.69 & 1.53 & 2.93 & 0.1 \\
\enddata
\tablenotetext{a}{Mass loss rates are given for the time period of interest.
For the isotropic case, the average mass loss rate of 5 $\times$ 10$^{-6}$
M$_{\sun}$ yr$^{-1}$ is used.}

\end{deluxetable}

 \begin{deluxetable}{lcccc}
 \tablecolumns{5}
 \tablewidth{0pc}
 \tablecaption{{\tt ChN} Dynamical and Spectral Results at t$_{\mathrm{SNR}}$ 400 yr \label{tab:results}}
 \tablehead{
 \colhead{Model} & \colhead{R$_{\mathrm{FS}}$} & \colhead{M$_{\mathrm{ej}}$} & 
 \colhead{M$_{\mathrm{swept~up}}$} &  \colhead{He-like Fe} \\
 \colhead{} & \colhead{pc} & \colhead{M$_{\sun}$} & \colhead{M$_{\sun}$} & \colhead{keV}}
 \startdata
 m15Iso & 1.99 & 2.0 & 0.6 & 6.662 \\
 m15C   & 1.62 & 6.0 & 2.4 & 6.665 \\
 m15O   & 1.98 & 5.9 & 8.6 & 6.676 \\
 \enddata
\end{deluxetable}

\end{document}